# Audit fees in auditor switching

Sarit Agami, Department of Economics, The Hebrew University of Jerusalem, Israel

**Abstract**

The auditor work is examining that a company's financial statements faithfully reflect its financial situation. His wage, the audit fees, are not fixed among all companies, but can be affected by the financial and structural characteristics of the company, as well as the characteristics of the firm he belongs to. Another factor that may affect his wage is an auditor switching, which can be resulted from changes in the company that may influence the fees.

This paper examines the effect nature of the auditor switching on his wage, and the factors of the company characteristics and the economy data which determine the wage at switching. A product of the research are tools for predicting and evaluating the auditor wage at switching. These tools are important for the auditor himself, but also for the company manager to correctly determine the wage due to the possibility that the quality of the audit work depends on its fees.

Two main results are obtained. First, the direction of the wage change in the switching year depends on the economic stability of the economy. Second, the switching effect on the direction and the change size in wage depends on the change size in the company characteristics before and after switching – a large change versus a stable one. We get that forecasting the change size in wage for companies with a large change in their characteristics is paralleled to forecasting a wage increasing. And vice versa, forecasting the change size in wage for companies with a stable change in their characteristics is paralleled to forecasting a wage decreasing. But, whereas the former can be achieved based on the company characteristics and macroeconomic factors, the predictability of these characteristics and factors is negligible for the letter.

## 1. Introduction

The role of auditor is to examine that the financial statements of a company adequately reflect the financial situation of the company, and that they do not contain a material misrepresentation. The audit fees, i.e., the auditor wage, is not uniform among all companies, but is determined in a negotiation between the auditor and the company.

It is interesting to examine whether there is any pattern and consistency in determining this wage. This is important for the auditor when deciding on an employment place, and for the company's management as well for determining the audit fees correctly due to the possibility that the quality of the audit work depends on its fees.

Many studies have discussed this issue, and examined the factors affecting the audit fees. Overall, the key determinants are divided into properties of the audited company- financial and structural, and properties of the audit. The financial properties of the audited company are finance risk (Kikhia, HY (2015), Castro et al. (2015), Kimeli, EK (2016), Mohammed et al. (2018)) and profitability (Musah, A. (2017), Super et al., (2019), Kimeli, EK (2016), Naser et al., 2016)), where the structural properties are the company size ( Kikhia, HY (2015), Musah, A. (2017), Kimeli , EK (2016), Al- mutairi , 2017 Super et al. (2019), Mohammed et al. (2018)) and complexity (Kimeli , EK (2016), Mohammed et al. (2018), Urhoghide et al. (2015), Castro et al. (2015) ,Naser et al., 2016), the industry type (Liu S. (2017)), and the structure of CEO and directors ( Harjoto, MA, Laksmana , I., & Lee, R. (2015)). The properties of the audit are tenure (Kimeli, EK (2016), Urhoghide et al. (2015)), age, gender, and educational background (Liu, S. (2017)), timing (Mohammed et al. (2018)), and affiliation (big four firms) (Musah, A. (2017), Kimeli, EK (2016), Mohammed et al. (2018), Al- mutairi (2017), Castro et al (2015), Horowitz et al.(2017)).

The direction of the relationship between some of these factors and the audit fees is ambivalent over the studies, probably due to dependence on the regulatory framework at the examined country and the type of companies considered in each study.

Auditor switching can also have an effect on the auditor wage, and according to the existing literature, the effect is a decrease in wage in the year immediately following the switching: Simon Francis (1988) examined data from American public companies that switched an audit firm in the years 1979-1984, and found that there was a significant decrease in fees in the year immediately after the switching. The decrease moderated in the second and third year after the switching and only in the fourth year did the auditor's fees return to its pre-switching level; Weiss, D. et al.

(2021) studied data from public companies in Israel that changed audit firm between 1/1/12-3/31/18 and found a decrease in fees. In the data they have examined, the auditor's report in the year preceding the switching was not drafted in the uniform version for over half of the companies, partly due to the "going concern" note and the attention paragraph. Castro et al. (2015) found that in the case of auditor switching, the large clients paid less in the first year of the auditor's employment.

Generally, a decrease in auditor's wage in the first year of his employment known as lowballing (DeAngelo, LE (1981)). But the framework of auditor switching is different because it can be resulted from changes in the company that may affect the fees. The changes include an appointment of a senior position (Firth, M. (1999)), Going concern note (Carcello, J. V., & Neal, T. L. (2003)), restating the financial statements, whether the restating is due to a mistake or a deliberate omission (Hennes, K. M., et al. (2010)), and accelerated growth or a wide geographical spread of the company (Johnson, W. B., & Lys, T. (1990)).

In this paper we will examine the effect of auditor switching on his hourly wage, and in particular the factors that determine the wage at switching. The goal is to provide tools for forecasting and evaluating the auditor wage at switching, which will be used by both the auditor and the company manager during wage negotiations. The paper considers public companies in Israel that have switched auditors in the years 2007-2012. This range of years was chosen because it includes a time of high tide in the economy (year 2007), low tide in the economy - economic crisis (year 2008), and a gradual exit from the crisis in 2009-2012. Only one year was considered before the crisis, because only starting in 2007 were the public companies in Israel required by law to report the audit fees as well as the number of hours spent on the audit work (Companies Law). This range of years differs, for example, from Weiss's paper because it is characterized by an economic depression relative to the years examined by Weiss's paper. The collection of factors we refer to in this paper includes financial and structural characteristics of the company, as well as macroeconomic factors. To capture the switching effect, we refer to the relative change in fees and in each of the aforementioned collection of factors in the year of the switching compared to the year before.

The paper is structured as follows: Section 2 describes the database, the statistical analysis of the data is described in Section 3, results and discussion are presented in Section 4, and Section 5 presents a brief summary.

## 2. The Database
### 2.1 Sources and Content

The data includes information on Israeli companies traded on the Tel-Aviv Stock Exchange (TASE) who switched an auditor in the years 2007-2012. The list of these companies according to the switching year was downloaded from the MAGNA website (Magna Web), and the annual report of each company was downloaded from the MAYA website (Maya Web).

From each report, the characteristics of the company at the switching year and at the year before the switching were taken, as follows:

A. Total wage and total hours for audit services and tax services.

B. Financial characteristics of the company

1) Total Assets (sum of total current assets and total non-current assets)

2) Total Liabilities (sum of current liabilities and non-current liabilities). Long-term liabilities were not considered.

3) Total Capital

4) Gross Total Revenue, which includes revenue from sales/services, financial income, and Other Income.

5) Total Net Profit

6) Basic Earnings per Share (EPS Basic) for Agora (0.01 of one NIS).

In case of distinguishing between net EPS Basic to EPS Diluted, the EPS Basic was considered. In addition, in case of information on EPS from continuing activities vs. EPS from discontinuing activities, the former was considered.

7) Cash and Cash Equivalents at the beginning of the financial year.

8) Cash and Cash Equivalents at the end of the financial year.

The currency unit of these characteristics is NIS. In the case of a foreign currency unit in the report, the conversion was made to NIS according to the exchange rate as it was on December 31 of the referring year (Historical rates).

Hourly wage for audit services and tax is calculated by total wage divided by number of working hours.

For each of the characteristics 1)-8) as well as for "hourly wage", the relative change in the characteristic in the switching year compared to the year preceding the switching is calculated. That is, if we denote by X the variable in the switching year t, then the relative difference is (Xt-Xt-1)/Xt-1.

A value of zero in one of the financial characteristics was replaced by a value of 1 for the purpose of calculating the relative change, except for EPS where a value of zero was replaced with the value $10^{-10}$ .

C. Structural characteristics of the company

1) Sector classification in TASE (this data was taken from the MAYA website according to "company details").

This includes classification into Sector and Super-Sector.

The categories of a super-sector are high- technology, finance, Real, and Financial Instruments.

The categories names of a sector, and their abbreviations as they appear in the paper are:

| Sector Name | Biomed | Technology | Financial Services | trade and services | Real Estate and Construction | Industry | Investment and Holdings | Energy and oil and gas exploration | Financial Instruments |
|---|---|---|---|---|---|---|---|---|---|
| name in the paper | Biomed | Technology | FinanceServ | TradeServ | Real-Estate | Industry | Investment | Energy | FinanceInstru |

2) Characteristics that were considered in the switching year are:

i) Going concern note (yes/no)

ii) Appointment of Senior Position (yes/no)

iii) The type of change in the size of the audit firm in the switching year compared to the year before. This includes a change from small to large, large to small, no change (i.e., small to small, or large to large). An audit firm is considered as "big" If she is one of the "big five" audit firms (Big 5), which include the Kost Forer Gabbay & Kasierer, KPMG, PwC, Deloitte, BDO. Otherwise considered "small" audit firm.

Remarks:

(i). For companies that had several auditors, including in abroad or of subsidiaries abroad, the wage and hours of the audit in Israel were considered, but the relevant characteristics were taken from the consolidated report in the annual report.

(ii). For a company with subsidiaries, only one of these companies was taken into account to avoid duplication of information (For example, Lito Real Estate Ltd. vs.

Lito Group Ltd., or Excellence which had a number of sub-companies such as Excellence Tavor, Excellence Adir, Atzmon Bonds, etc.)

(iii) Companies for which the distribution of wages and hours between partners of several auditors was not clear, were removed from the analysis (this situation could happen when the company had subsidiaries abroad, or in the case where the old and new auditors worked together before the switching).

(iv). Companies that had no report on Maya website or had a quarterly report rather than an annual report since the company was deleted from TASE, were removed from the analysis.

(v). In case of a very high and unusual hourly wage for some company, the wage was considered as is without any trimming.

In addition to these characteristics, macroeconomic factors in Israel were examined. They were taken from the CBS website and include the consumer price index in Israel (CBS_1), average wage in the economy (CBS_2), GDP (CBS_3), consumption (CBS_4). These factors were examined at the switching year and the year before, and for each of them the relative change was calculated, as defined above.

Abbreviated names in the paper for the relative difference in each of the above measured characteristics are as follows:

| Feature | Total Assets | Total Liabilities | Total Capital | Gross Total Revenue | Total Net Profit | Basic Earnings Per Share | Cash and Cash Equivalents at the beginning of the financial year | Cash and Cash Equivalents at the end of the financial year |
|---|---|---|---|---|---|---|---|---|
| name in the paper | Assets | Liabilities | Capital | Gross Revenue | Net Profit | EPS Basic | Cash begins | Cash end |
| Feature | Going concern note | Appointment of Senior Position | Type of size change of audit firm | Hourly wage | GDP | | | |

| name in the paper | Going concern | Senior Position | Auditor Type | wages | GDP | Market Wage | Price | |
|---|---|---|---|---|---|---|---|---|

## 2.2 Missing Data

The dataset contained several companies with missing EPS. The distribution of these companies each year and according to sector is described in Table 1. Since these companies make up a relatively small percentage of the data collection, missing EPS was imputed. The imputation was done by a regression model of EPS on a financial characteristic which found to be correlated with the highest each year. All the analysis in this paper is based on the collection that includes imputation of deficiencies in the EPS.

| year | Number of companies | Sector |
|---|---|---|
| 2007 | 1 | Real-Estate |
| 2008 | 4 | Real Estate (2), Investment (2) |
| 2009 | 5 | Real Estate (2), Investment (2), FinanceServ (1) |
| 2010 | 3 | Technology (1), FinanceInstru (1), Investment (1) |
| 2011 | 5 | Real Estate (1), FinanceInstru (2), Investment (2) |
| 2012 | 2 | Real Estate (1), FinanceInstru (1) |

**Table 1.** The distribution companies with a missing value in EPS, and their affiliation to the sector. In the sector column, the number of companies in each sector is described in parentheses.

## 3.2 Characteristics of the companies

According to the construction of the database, a collection of 143 companies was obtained. The distribution of the number of companies in each year and the distribution of companies according to affiliation to "super sector" and "sector " in each year (relative frequency) is described in Table 2. It can be seen that in this data, there are on average 20 companies each year that switched an auditor. This except the year 2008 when the highest number of auditors was switched (the number of companies that switched an auditor is almost twice that of the other years), as we would expect due to a year of economic crisis. The auditor switching was mainly common among companies in the real super-sector, this is the main majority, especially in the real construction sector. About 30% of the companies were in the

high - tech super-sector, mainly in the "technology" sector, and the auditor switching was minor in the financial instruments sector. The distribution of the companies by the size of audit firm is described in Table 3, and the distribution (relative frequency) of having a going concern note and appointing a senior position in the switching year is described in Table 4. We see that in the switching year compared to the previous year, in slightly more than 50% of the companies there was no difference in the size of audit firm, 20% changed from a large to a small firm, and the rest of the companies changed from a small to a large firm. Specific to the year, 2009 is characterized by the lowest percentage of changing audit firm from small to large, and by the highest percentage of no difference in the audit firm. The highest percentage of changing from small to large audit firm was done in 2007 (before the economic crisis) and in 2011 (the year of exit from the economic crisis). Likewise, in the switch year, there was a going concern note for 30% of the companies, and in 78% of the companies there was an appointment of a senior position. An upward trend in the appointment of a senior position began in 2008, and the relative frequency of the highest going concern note was in 2009 - a year after the crisis, and in 2012 - a year out of the crisis.

A comparison of the companies in the real super-sector (96 companies) versus the high-tech super-sector (38 companies) is described in Table 5. In the real companies relative to the high tech there was a higher percentage of companies where there was no difference in the size of audit firm, but a lower percentage for companies that changed from a large audit firm to a small one.

Also, a higher percentage of going concern note and appointment of a senior position was found in companies in the high-tech super-sector compared to those in the real super-sector.

|      |     | Super Sector | | | | | | | | | | |
|------|-----|--------------|---|---|---|---|---|---|---|---|---|---|
|      |     | High-tech | | | Finance | | Real | | | | | Financial Instruments |
| Year | n | Biomed | Technology | All | Financial Services | Trade and Services | Real Estate and Construction | Industry | Investment and Holdings | Energy and oil and gas exploration | All | Financial Instruments |
| 2007 | 20  |      | 0.30 | 0.30 |      |      | 0.30 | 0.20 | 0.10 | 0.05 | 0.65 | 0.05 |
| 2008 | 39  | 0.08 | 0.10 | 0.18 |      | 0.28 | 0.28 | 0.08 | 0.18 |      | 0.82 |      |
| 2009 | 20  | 0.10 | 0.15 | 0.25 | 0.05 | 0.05 | 0.25 | 0.20 | 0.20 |      | 0.70 |      |
| 2010 | 24  | 0.17 | 0.17 | 0.33 | 0.04 | 0.04 | 0.12 | 0.08 | 0.29 | 0.04 | 0.58 | 0.04 |
| 2011 | 20  |      | 0.20 | 0.2  |      | 0.15 | 0.15 | 0.05 | 0.20 | 0.15 | 0.7  | 0.10 |
| 2012 | 20  | 0.15 | 0.25 | 0.40 | 0.10 | 0.05 | 0.25 | 0.10 |      | 0.05 | 0.45 | 0.05 |
| **All** | **143** | **0.08** | **0.18** | **0.27** | **0.03** | **0.12** | **0.23** | **0.11** | **0.17** | **0.04** | **0.67** | **0.03** |

**Table 2.** Distribution of companies (relative frequency) by super-industry and industry in each year. The All columns describe the relative frequency of the total number of companies in each sector per year. The All row describes the relative frequency of the companies in each super-sector and in each sector beyond over all the years together. The distribution of the 143 companies according to their number in each year is described by the column titled n.

|  | Change in Size of Audit firm | | |
|---|---|---|---|
| year | No difference | Small to big5 | Big5 to small |
| 2007 | 0.4 | 0.4 | 0.2 |
| 2008 | 0.56 | 0.23 | 0.21 |
| 2009 | 0.65 | 0.10 | 0.25 |
| 20 10 | 0.50 | 0.29 | 0.21 |
| 2011 | 0.5 | 0.4 | 0.1 |
| 2012 | 0.55 | 0.20 | 0.25 |
| **ALL** | **0.53** | **0.27** | **0.20** |

**Table 3.** Distribution of companies (relative frequency) in each year by type of audit firm in the switching year in relation to the year before.

| year | Going concern | Senior Position |
|---|---|---|
| 2007 | 0.1 | 0.7 |
| 2008 | 0.33 | 0.79 |
| 2009 | 0.4 | 0.75 |
| 2010 | 0.25 | 0.79 |
| 2011 | 0.25 | 0.75 |
| 2012 | 0.45 | 0.85 |
| **ALL** | **0.3** | **0.78** |

**Table 4.** Distribution of companies (relative frequency) in each year according to holding a going concern note and according to holding a senior position appointment in each switching year.

|  |  | High-tech | Real |
|---|---|---|---|
| Auditor type | No difference | 0.39 | 0.58 |
|  | Small to big5 | 0.24 | 0.28 |
|  | Big5 to small | 0.37 | 0.14 |
| Going Concern | Yes | 0.39 | 0.28 |
| Senior Position | Yes | 0.84 | 0.75 |

**Table 5.** Distribution of companies (relative frequency) in the real super-sector and the high-tech super-sector according to the type of auditor switching, as well as according to having a going concern note and a senior position appointed in the switching year, years 2007-2012.

The distributions of the relative difference in the switching year compared to the year before the switching in each of the financial characteristics of the companies, over all the years together, are described in Figure 1. It can be seen that there are many outliers resulting from a large change in a certain characteristic between the two consecutive years (for example, a company with zero cash before the switching, and has a high positive value of cash in the switching year). Therefore, the analysis below will distinguish between the group of companies with such outliers, and the group of other companies which have reasonable relative differences.

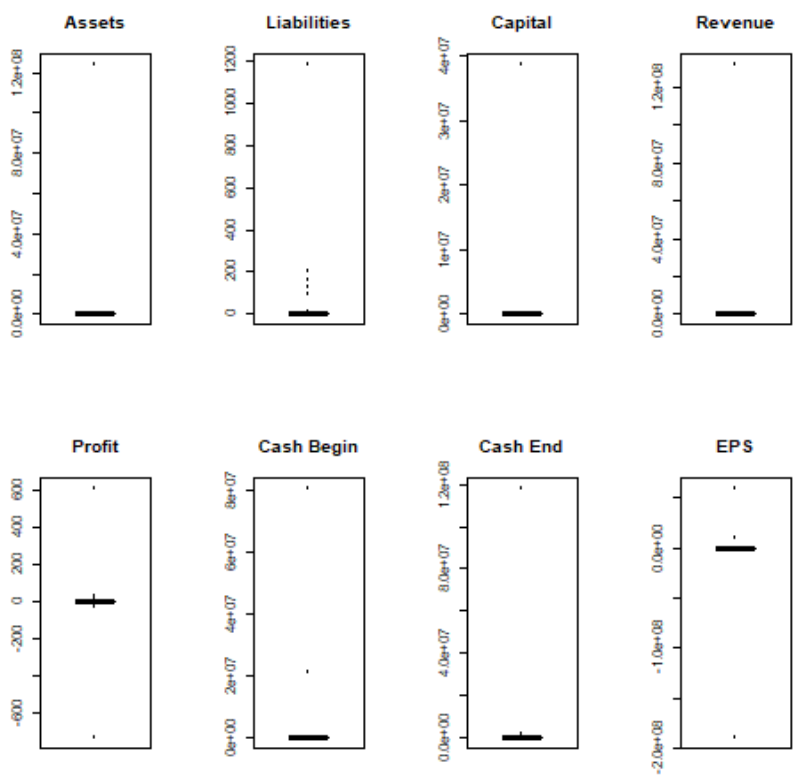

**Figure 1.** Distribution of relative difference of each financial characteristics, over all companies, years 2007-2012.

The correlations between the financial characteristics over the all years together are described in Figure A1 in Appendix A. It is possible to see the existence of a correlation between the financial characteristics, and for some there is a high correlation.

## 3.2 Change in hourly wage

The distribution of the relative change in hourly wage (hereinafter referred to as the "change in wage") per year is described in Figure 2. We see that the distribution includes both a negative relative change, meaning that wage after the switching is lower than that before the switching, and also a positive relative change, meaning that wage after the switching is higher than before the switching. A closer examination of the percentage of companies with a negative relative wage change versus a positive relative wage change is presented in Table 6. In general, in a little more than half of the companies (55%) the wage after the switching is lower than before the switching. An upward trend is seen from 2010 on in the percentage of companies in which the wage relatively lower at switching. However, the year 2009 (a year after the economic crisis in 2008) stands out in that for a large majority of companies the hourly wage in the switching year increased compared to the year before the switching. In a comparison between companies in the real super-sector versus companies in the high-tech super-sector over the all years together, as shown in Figure 3, the distribution of change in wages among these two sectors is quite similar, but there are several companies in the real super-sector that have a very large positive change compared to the on high-tech.

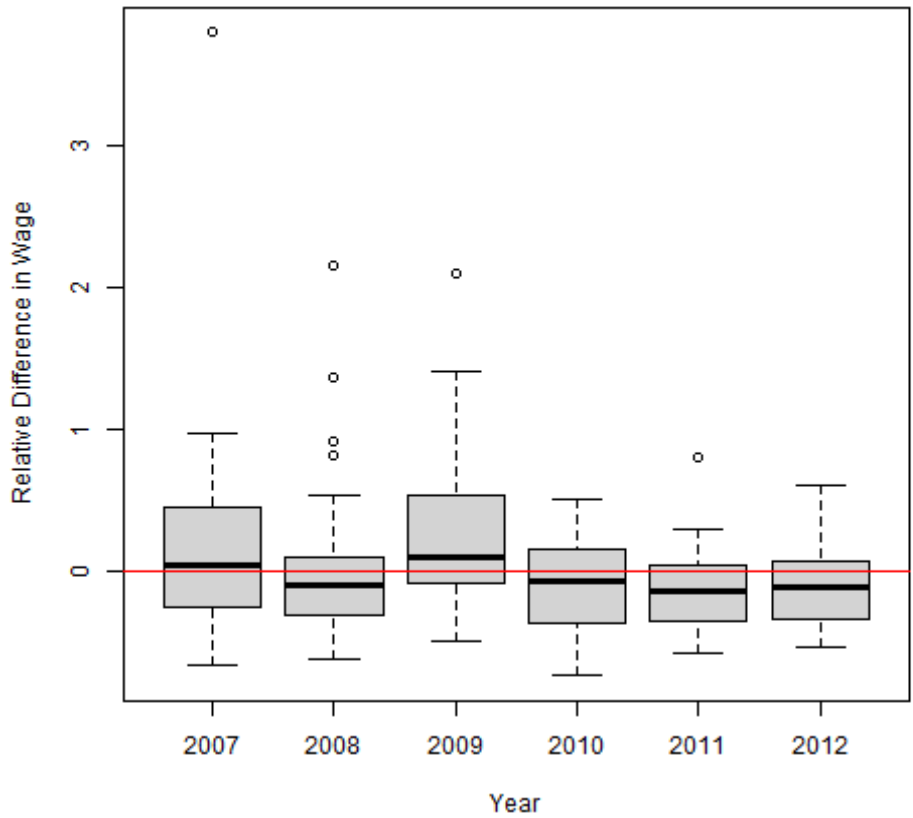

**Figure 2.** The distribution of the relative change in the hourly wage (wage in the switching year relative to the year before the switching) per year. A reference line is at zero height, so above it wages rose in the switching year, and vice versa.

|      | Relative difference in wages | |
| --- | --- | --- |
| year | $\Delta W < 0$ | $\Delta W > 0$ |
| 2007 | 0.45 | 0.55 |
| 2008 | 0.59 | 0.41 |
| 2009 | 0.3  | 0.7  |
| 2010 | 0.54 | 0.46 |
| 2011 | 0.65 | 0.35 |
| 2012 | 0.7  | 0.3  |
| **ALL** | **0.55** | **0.45** |

**Table 6.** Distribution of companies (relative frequency) each year according to the direction of the relative change in wage $\Delta W$, positive or negative.

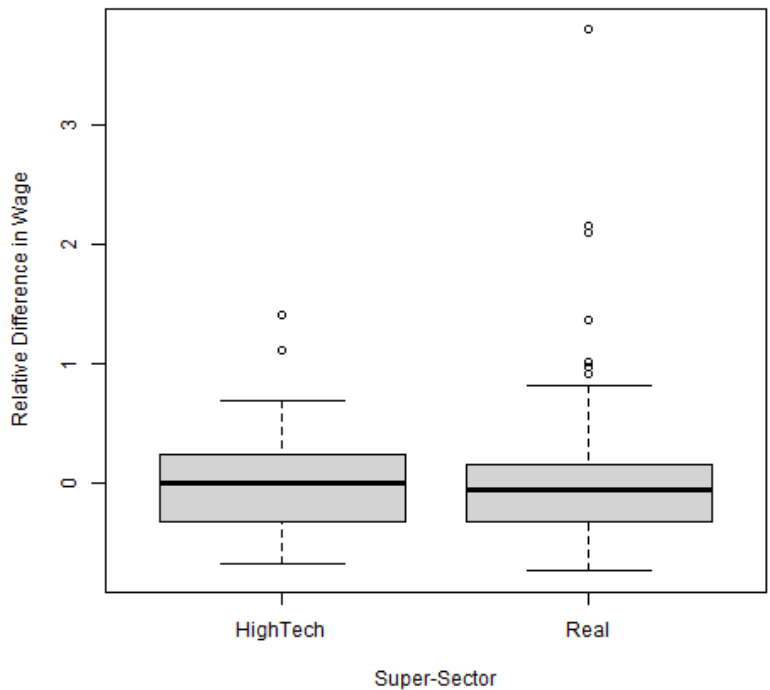

**Figure 3.** Distribution of the relative change in hourly wage in real super-sector companies versus high tech super-sector companies, over the years 2007-2012.

The correlations between the relative change in each of the financial characteristics and the relative change in wage over the all years together are described in Figure A1 in Appendix A. These correlations per year are described in Table A1 in Appendix A. It can be seen that the characteristics that correlate highly with the change in wage over the all years together are the relative change in assets, liabilities, and cash at the end of a period. In particular, these relationships exist in the real sector, but in the high-tech sector there are low correlations of financial characteristics with a relative change in wage.

Specifically for year and over the all companies together, high correlations are observed mainly in the years 2007-2009 (years around the economic crisis). The relative changes in financial characteristics that are highly correlated with the relative change in hourly wage, by year, are: 2008 (crisis year) - profit, 2009 (year immediately after the crisis) - assets, income, cash at the end of the period , 2010 - income, 2011 - share price , cash at the beginning of the period, year 2012 - liabilities.

In terms of structural characteristics of the companies, based on a Chi-square test for independence between categorical variables, there is no statistical relationship between the direction of the change in wage (positive/negative direction) and the type of change in the size of the audit firm. Furthermore, there is no statistical relationship between the direction of the change and the existence of a going concern note and the appointment of a senior position in the year of the change. A joint distribution of the direction of the change in wage and the sector for all companies together is shown in Table A2 in Appendix A. It can be seen that the proportion of companies with an increased wage is similar to the one with a decreased wage, except of investHold for which there is a higher proportion of companies with a decrease in wage compared to those with an increase, and there is also a tendency for an increase in wage in biomed. That is, in general, it seems that sector is not very coordinated with the direction of the change in wage.

**4.2 Distinguish between companies' types**

As mentioned above, there is great variability in the relative change in financial characteristics of the companies. Therefore, we will distinguish between companies with an outlier relative change in one or more financial characteristics (hereafter the "outlier companies") and companies with reasonable relative changes (hereafter the "regular companies"). The specific determination of an outlier relative change in any financial characteristic was made according to the Tukey method for outliers, as follows: Let X be some financial characteristic, and let $\Delta X$ the relative change of X. Let denote by IQR the inter-quarterly range of $\Delta X$, and by $Q_1, Q_3$ the lower and upper quarter, respectively. Then, a specific value $x^*$ of $\Delta X$ is considered an outlier if $x^* < Q_1 - 3IQR$ or if $x^* > Q_3 + 3IQR$. According to this criterion, we get 64 companies with outlier relative differences, and 79 companies with reasonable relative differences. We will examine the characteristics of each of these 2 groups.

2.4.**1 Outlier Companies**

Among the 64 outlier companies, 69% of them operate in the real super-sector, mainly in the realConstruction sector, 31% in the high-tech super-sector, and none of them is associated with the financial super-sector. Also, among these companies, 62% of them did not change the size of audit firm, 22% of them changed from a small firm to a large firm, and 16% from small to large. In the switching year, 44% of these companies had a going concern note, and 77% appointed a senior position in the company. Companies with a high percentage of a going concern note and an appointment of a senior position in the switching year are mainly concentrated in the years 2008, 2012. In terms of the direction of the relative difference in wage, in a little over half of them (53%) the relative wage decreased, and in the rest the relative wage increased. The main increase in wage was in 2009, that is, immediately after the year of the economic crisis, when in the next years the relative wage decreased in most of the companies. Details of these distributions by year are presented in Tables B1-B4 in Appendix B. The correlation of relative difference in each of the financial characteristics and wage are described in Figure B1 in Appendix B. It can be seen that there are differences in the financial characteristics that have a high correlation between them. Also, a relative difference in wage is highly correlated with a relative difference in assets, liabilities, and cash at the end of a period.

By creating an indicator for a wage difference smaller than zero or larger than zero, it is obtained from a Chi- squared test for independence that there is no relationship between this indicator and each of the structural characteristics of size of audit firm switching, going concern note, and senior position appointment. It was not possible to apply this test for a super-sector and sector, due to the existence of a very low frequency in some categories.

### 2.4.2 Regular Companies

Among the 79 regular companies, 66% of them operate in the real sector, 23% in the high-tech sector, 5% in the financial instruments sector, and 6% in the financial sector. Also, among these companies, 46% of them did not change the size of the audit firm, 30% changed from a small firm to a large one, and 24% from a small firm to a large one. In the switching year, 19% of them have a going concern note, and 78% appointed a senior position in the company. In terms of direction of the relative difference in wage, the relative wage decreased for 56% of them, that is, a little more than half, and vice versa for the other companies. Details of these distributions by year are presented in Tables C1- C4 in Appendix C.

Correlations of relative difference in each financial characteristic and wage are described in Figure C1 in Appendix C.

It can be seen that there are financial characteristics that have a high correlation between them, but the correlations with the wage difference are very low.

From a Chi- squared test for independence, it is obtained that there is no relationship between an indicator of a direction of the wage difference and any of the structural characteristics of size of audit firm switching, going concern note, and appointment of a senior position. It was not possible to apply this test for a super-sector and a sector, because the existence of very low incidences in several categories.

### 2.4.3 Comparison of Outlier companies vs regular companies

The percentage of outlier companies who did not change the size of audit firm is higher compared to the regular companies. But in the regular companies a higher percentage of companies that changed from a small to a large or a large to a small audit firm. Also, there is a higher percentage of outlier companies with a going concern note compared to the regular companies, but the percentage of companies that have appointed a senior position is similar among the outlier and regular companies. The effect of the economic crisis in 2008 is higher on the outlier companies compared to the regular. The years 2008, 2012 have a higher percentage of going concern note in the outlier companies, but in the regular companies there is no similar effect of the crisis year on the going concern note.

According to all of the above, the companies with large fluctuations compared to those with small fluctuations can be characterized as more active in the high-tech sector and do not operate in the financial sector, having a higher percentage (more than 2 times) of going concern note, a slightly higher percentage of type of audit firm size, and a higher variability in the relative difference in wage (a higher relative difference compared to the companies with the small fluctuations) as can be seen in Figure 4.

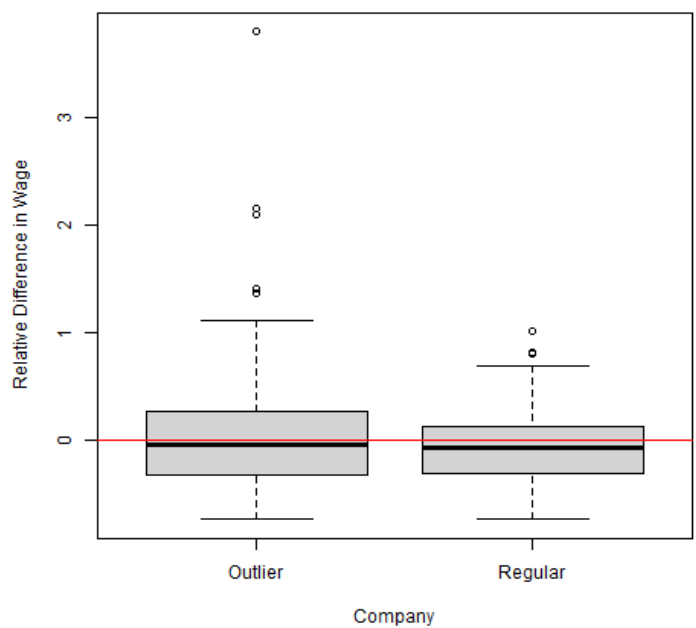

**Figure 4.** The distribution of the relative difference in hourly wage among companies with large fluctuations in their characteristics versus small fluctuations. A reference line at zero height, above it a higher relative wage in the year of the exchange, and vice versa.

## 3. Modelling the relative wage
### 3.1. Modelling the difference in wage

We consider the modelling of the difference in wage according to the relative difference in each of the company's characteristics, financial and structural, as well as according to each of the macroeconomic factors. Since the difference in wage includes positive and negative values, depending on the direction of the wage change in the year of the exchange, this modelling describes some kind of wage change (and not a specific positive or negative direction), and explains the size of the wage change.

Due to high correlations in the relative differences among some of the financial characteristics of the companies, as mentioned above, a linear regression model will not fit here (multicollinearity problem, as can be seen in the calculation of VIF index in Appendix D, when macroeconomic characteristics are also added). As an alternative, one can use principal component regression (PCR) based on principal components analysis (PCA) (Greenacre (2022)). In general, the PCA method is an exploratory tool. It produces a system of new variables called principal components, based on original variables. Each of the principal components is a linear combination of the original variables. Such a combination is constructed in a way that captures the highest variance among all the original variables, and such that the principal components are uncorrelated with each other. The PCR method (Jolliffe (2010)) is a regression model of an explained variable on the above principal components that are explanatory variables in the model. One way to choose the number of the principal components for the regression model is according to a high percentage of cumulative explained variance.

In our framework, we create principal components based on the characteristics of the company and the macroeconomic factors, choose the number of the principal components to be such that produces about 80% cumulative explained variance, and then we apply a linear regression model of the relative change in wage on these principal components.

We perform the PCR separately for outlier and regular companies, over all the years together.

In addition, to examine the year effect, we fit the PCR for each year separately, over the all companies. In examining this effect, we do not make a separation into outlier and regular companies, due to the small number of companies each year. Although the sample size is still small when referring to all the companies per year, so it is less possible to apply a statistical inference from a model, still we can try to learn about the differences if they exist. First we perform the modelling based on relative differences of the company's characteristics only, and then we add the relative differences in the macroeconomic factors.

All calculations were done by R software, using the *PCAmixed* and *lm* procedures, when the letter procedure was applied to the results of the former procedure.

Each principal component is characterized by the 'loadings', which are the correlations between the original variables and the given principal component. It is acceptable to refer to a loading threshold of 0.4 (Williams (2010)) Therefore, we will characterize each principal component according to such a threshold value. In the case where the loadings are lower than 0.4, each principal component will be characterized by the variables with the highest loading.

The correlations are used to determine which variables are important for interpreting a principal component.

**3.1.1 Modelling wage by company type, years 2007-2012**

Checking the conformity of the outlier and regular companies data to PCA as expressed by KMO (Kaiser (1974)) and Bartlett tests (Bartlett (1951)), is presented in Tables 8,9. The PCA and PCR results are also included in these tables. The fit to PCA is good for each of the above 2 groups.

For the outlier companies, a PCA fit to yield 9 principal components with a cumulative explained variance of 82%. The first principal component, which accounts for about 17% of explained variance, is mainly affected by a relative change in assets, liabilities, cash at the end of a period, with a positive weight of each of them on the principal component; A second principal component, which accounts for about 13% of explained variance, is mainly affected by a relative change in the stock, net capital, as well as by sector. The other principal components are mainly influenced by sector. A linear regression of the change in wage on these 9 principal components produces the components 1, 4, 5 as significant, with the coefficients $\beta_{PC\#1}$ = 0.24, $\beta_{PC\#4}$ = 0.12, $\beta_{PC\#5}$ = -0.16, and with $R^2_{adj}$ =0.40.

For a given principal component, the sign of the regression coefficient together with the weight sign of the main loadings yield that the relative change in assets, liabilities, and cash at the end of a period have a positive effect on the relative change in wage, that is, a greater relative change in one of them leads to a greater relative change in the wage. This is compared to a sector that has both a positive and negative effect.

A more in-depth look at the relation sector-change in wage is presented in Figure 5 and Table 7. Table 7 describes the sector distribution against the direction of the change in wage according to the type of company – outlier or regular. The distribution is over the total number of companies (143). Its purpose is to examine which sector is worth working with for the auditor in the switching year. This is compared to Figure 5 which describes the distribution per sector (and not over the all sectors) for the outlier companies only. By the variation presented in this figure, it is possible to learn which sector was more affected at the switching year in terms of the change in wage. It can be seen an upward change in wage in Industry, positive and negative change in the other sectors, while in the Energy, Investment, and the Trade Services sector the main change in wage is negative, meaning a small wage after the switching. The size of the change (with no specific direction) is greater (with greater variability) in the Technology sector.

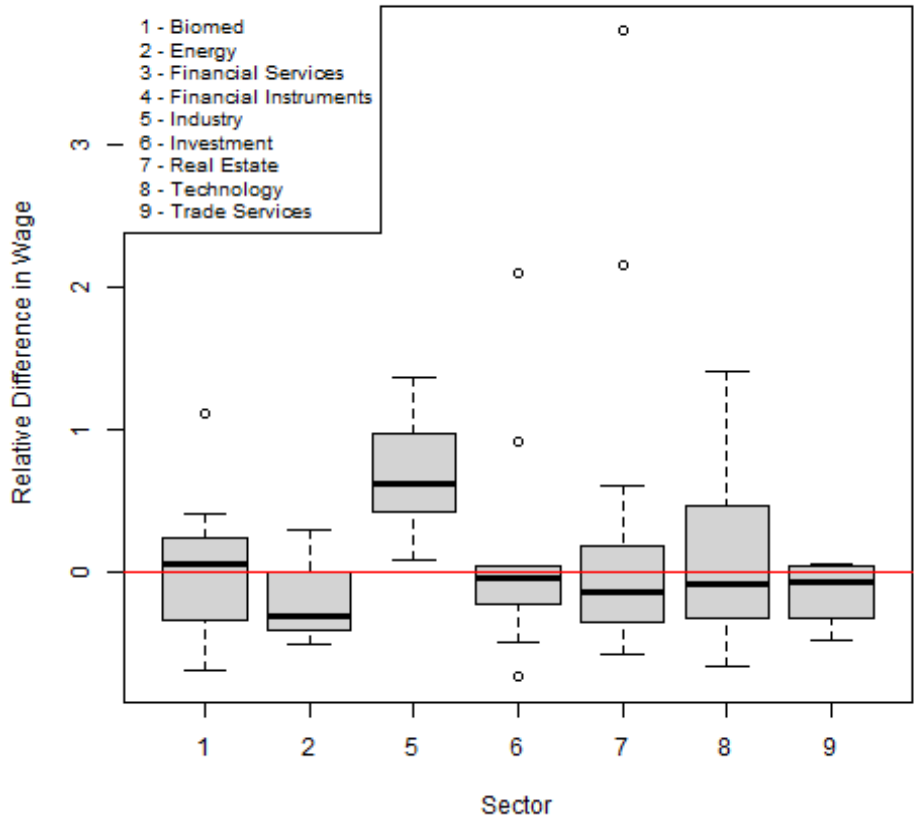

**Figure 5.** The distribution of the change in the relative wage by sector among the outlier companies, years 2007-2012. A reference line It is at zero, meaning a situation where there is no change in wage after the switching compared to before.

|  | outliers |  | regular |  |
| --- | --- | --- | --- | --- |
|  | ΔW <0 | ΔW >0 | ΔW <0 | ΔW >0 |
| biomed | 0.06 | 0.09 | 0.00 | 0.03 |
| EnergyGas | 0.05 | 0.02 | 0.03 | 0.00 |
| FinanceService | 0.00 | 0.00 | 0.04 | 0.01 |
| FinanceTools | 0.00 | 0.00 | 0.04 | 0.03 |
| industry | 0.00 | 0.08 | 0.09 | 0.05 |
| InvesHold | 0.11 | 0.06 | 0.11 | 0.05 |
| RealConstruction | 0.17 | 0.14 | 0.08 | 0.09 |
| technology | 0.11 | 0.05 | 0.10 | 0.10 |
| TradeService | 0.03 | 0.03 | 0.08 | 0.09 |

**Table 7.** Sector distribution against the direction of the change in wage according to the type of company – outlier or regular (the distribution is per company and expresses relative frequency out of the total number of outlier companies):

By adding a relative change in macroeconomic factors to the analysis, the first 10 PCs produce 82% cumulative explained variance. The first and third PCs are similar to the aforementioned PCs 1 and 2, but the second PC includes a relative change in GDP and the sector. The other PCs mainly include a sector. A linear regression on these 10 PCs produces the components 1,2,4,8 as significant. The regression coefficients are

$\beta_{PC\#1}$ = 0.15, $\beta_{PC\#2}$ = -0.20, $\beta_{PC\#4}$ = -0.13, $\beta_{PC\#8}$ = -0.16, and the goodness of fit is $R^2_{adj}$ =0.43.

As before, a relative change in assets, liabilities, and cash at the end of a period, have a positive effect on the relative change in wages and a mixed effect of an industry (as expressed by the negative sign of the coefficient of the second component and the positive sign of the coefficients of components 4,8). In addition to the change in GDP, there is a negative effect on the change in wages, meaning that a relatively large change in GDP leads to a smaller relative change in wages. The percentage of explanation of the relative change in wages according to their main components in the second model is higher and is 43%.

For the non-outlier companies, PCA adjustment to the data produces 12 main components with a cumulative explained variance of 82%, where a linear regression of the change in wages on these main components produces only the 11th component that is affected by industry as significant at a significance level of 7%, where $\beta_{PC\#11}$=0.07, and $R^2_{adj}$=0.03.

By adding macroeconomic characteristics to the analysis, the first 12 main components produce 79% cumulative explained variance, and a linear regression of the change in wages on these components produces only the 12th component as significant (which is affected by industry), where $\beta_{PC\#12}$ = 0.08 and $R^2_{adj}$=0.04.

That is, the percentage of explanation of the relative change in wages according to the main components in both models is negligible (3%-4%).

### 3.1.2 Modelling wage for all companies together per year

Modelling the relative change in wages for each individual year is done based on the relative change in the financial characteristics of companies only. Relative change in macroeconomic factors were not taken into account because each of them is an annual measurement, therefore the value is the same for all companies in a given year. This modelling is presented in Appendix E, Tables E1-E2.

Examining the suitability of data that includes the financial characteristics of all companies to PCA is presented in Table E1. A good fit is seen for every year, except for the year 2009 where the fit is borderline (Bartlett not significant at all) . The PCR results are shown in Table E2.

The percentage of variance captured by the first principal component in years of economic stability is about 22%-24%. This percentage also exists in the year of the economic crisis (2008). That is, the effect of the crisis is not expressed in the PCA results, but it is expressed in the year immediately after The crisis, in which the first principal component occupies only about 16% explanation of the variance.

In terms of the composition of the main components: in 2007 (a year before the economic crisis), the first component is mainly affected by a relative change in assets, net capital, cash at the end of a period, liabilities, the second component is mainly affected by industry, and the following main components include reference to super industry and industry.

In the year of the crisis, 2008, the first component is affected by the same characteristics as in 2007 with the exception of an effect from a relative change in income instead of liabilities in the first component. The other components include a reference to the industry. In the year immediately following an economic crisis (2009), the first component is mainly affected by industry and super-industry, and the following components are mainly affected by industry. Two years after the crisis: the first component is mainly affected by a relative change in profit, share, net capital, as well as by industry, and the following components are mainly affected by industry. In the following years there is a similarity in terms of the first component to that of 2007, but the second component is mainly affected by a relative change in the stock, net capital or profit, and by industry, and the following components are mainly affected by industry.

To summary, the main components are mainly influenced by the financial characteristics of the companies as well as the affiliation of each company to a certain industry. Specifically, during a crisis, the industry is more important, but in the years around the crisis year, a change in the ratio of assets, liabilities, revenues, profit, and net worth is also important. Treating these main components as explanatory in a linear regression on the difference in wage (PCR), produces a relatively low explained variance percentage of the difference in wage (ranging between 6% (2008) and 15% (2011), except for 2007 (76%), but in 2010 and 2012, the model does not produce an explanatory result (all the main components are accepted as non-significant), but at the same time, it is important to remember that this is a relatively small number of observations, as well as a high volatility between the observations.

Table 8. Modeling the relative difference in wage in outlier companies, years 2007-2012

| OUTLIERS COMPANIES | | | | | | | | | | | | |
|---|---|---|---|---|---|---|---|---|---|---|---|---|
| | **Without economic factors** | | | | | | | | | | | |
| PCA | KMO=0.56, p-value Bartlett= 1.76E-160 | | | | | | | | | | | |
| | | PC1 | PC2 | PC3 | PC4 | PC5 | PC6 | PC7 | PC8 | PC9 | PC10 | TOTAL |
| | variance( %) | 16.72 | 13.20 | 11.75 | 8.60 | 7.87 | 7.40 | 6.11 | 5.78 | 4.77 | | 82.20 |
| | Main loadings (weight *) | Assets (0.92) Liabilities (0.90) Cash end (0.92) | EPS (0.65) Capital (0.68) Sector (0.66) | Super-Sector (0.67) Sector (0.79) | sector (0.44) | sector (0.40) | Cash Begin (0.74) | Sector (0.78) | sector (0.80) | Sector (0.62) | | |
| PCR | Linear regression of $\Delta W$ on PCs: PCs #1, 4, 5, are significant. Keeping these PCs in the regression yields $\beta_{PC\#1}$** = 0.24, $\beta_{PC\#4}$ = 0.12, $\beta_{PC\#5}$ = -0.16, $R^2_{adj}$***=0.40 | | | | | | | | | | | |
| | **With economic factors** | | | | | | | | | | | |
| PCA | KMO=0.51, p-value Bartlett = 2.06E-204 | | | | | | | | | | | |
| | variance (%) | 15.11 | 11.34 | 10.98 | 9.69 | 7.29 | 6.82 | 6.11 | 5.72 | 4.68 | 4.26 | 81.98 |
| | Main loadings (weight) | Assets (0.70) Liabilities (0.68) Cash End (0.70) | GDP (0.42) Sector (0.49) | EPS (0.71) Capital (0.74) Sector (0.62) | Sector (0.41) | Sector (0.43) | Senior Position (0.40) | Cash Begin (0.71) | Sector (0.69) | Sector (0.81) | Sector (0.61) | |
| PCR | Linear regression of $\Delta W$ on PCs: PCs #1,2, 4, 8 are significant. keeping these PCs in the regression yields $\beta_{PC\#1}$ = 0.15, $\beta_{PC\#2}$ = -0.20, $\beta_{PC\#4}$ = -0.13, $\beta_{PC\#8}$ = -0.16, $R^2_{adj}$ =0.43. | | | | | | | | | | | |

* The characteristics that have the most influence on the PC are reported, where the correlation of each characteristic with the PC is reported in brackets. ** $\beta_{PC\#k}$ denotes the regression coefficient of the k-th PC *** $R^2_{adj}$ denotes the adjusted-$R^2$.

PCR is the linear regression of the relative difference in wage $\Delta W$ on the PCs.

Table 9. Modeling the relative difference in wage in regular companies, years 2007-2012

| REGULAR COMPANIES | | | | | | | | | | | | | | |
|---|---|---|---|---|---|---|---|---|---|---|---|---|---|---|
| | **Without economic factors** | | | | | | | | | | | | | |
| PCA | KMO=0.53, Bartlett p-value= 5.51E-21 | | | | | | | | | | | | | |
| | | PC1 | PC2 | PC3 | PC4 | PC5 | PC6 | PC7 | PC8 | PC9 | PC10 | PC11 | PC12 | TOTAL |
| | variance % | 11.49 | 10.25 | 8.81 | 8.74 | 8.00 | 6.23 | 5.70 | 5.51 | 5.00 | 4.67 | 4.36 | 3.56 | 82.31 |
| | Main loadings (weight*) | Assets (0.53) Liabilities (0.44) Sector (0.43) | EPS (0.43) Profit (0.40) Super-Sector (0.59) sector (0.7) | Super-Sector (0.56) Sector (0.68) | Super-Sector (0.82) sector (0.85) | Sector (0.57) | sector (0.60) | Sector (0.39) Senior Position (0.37) | Sector (0.59) | Sector (0.63) | sector (0.60) | Sector (0.51) | Cash Begin (0.28) sector (0.27) | |
| PCR | All PCs are insignificant at alpha=0.05, except PC #11 which is significant at 0.07. keeping this PC only in the regression yields $\beta_{PC\#11}$**= 0.07, with $R^2_{adj}$ ***= 0.03 | | | | | | | | | | | | | |
| | **With economic factors** | | | | | | | | | | | | | |
| PCA | KMO=0.47, Bartlett= 1.07E-92 | | | | | | | | | | | | | |
| | Variance % | 11.1 5 | 9.9 5 | 8.88 | 7.89 | 7.26 | 6.8 2 | 5.4 9 | 5.28 | 4.7 2 | 4.40 | 3.96 | 3.5 1 | 79.3 1 |
| | Main loadings (weight) | Price (0.46) | GDP (0.76) Market Wage ( 0.64) | EPS (0.44) Super-Sector (0.51) Sector (0.63) | Super-Sector (0.75) sector (0.85) | Super-Sector (0.45) Sector (0.62) | sector (0.46) | Sector (0.66) | Sector (0.66) | sector (0.54) | Senior Position (0.3) | sector (0.70) | sector (0.44) | |
| PCR | Only PC #12 is significant. Keeping this PC only in the regression yields $\beta_{PC\#12}$ = 0.08, with $R^2_{adj}$ =0.04 | | | | | | | | | | | | | |

\* The characteristics that have the most influence on the PC are reported, where the correlation of each characteristic with the PC is reported in brackets. ** $\beta_{PC\#k}$ denotes the regression coefficient of the k-th PC *** $R^2_{adj}$ denotes the adjusted-$R^2$.

PCR is the linear regression of the relative difference in wage $\Delta W$ on the PCs.

## 3.2. Modelling a specific direction of the wage difference

Dividing the companies according to the direction of the change in wage, that is, an increase or decrease in wage in the switching year compared to the previous year, there are 65 companies with an increase in wage, and 79 companies with a decrease in wage. To learn about factors that affect the size of the wage change in each direction, a PCR model was fitted in each of these 2 groups. The results are presented in Tables 10-11.

For the companies that experienced an increase in wage, a model based on the characteristics of the companies (financial and structural) produced the components PC1 and PC7 as significant, with the regression coefficients $\beta_{PC\#1}$ = 0.22, $\beta_{PC\#7}$ = -0.12, and with $R^2_{adj}$ =0.48.

PC1 is mainly affected by a relative change in assets, liabilities, and cash at the end of a period. Each of them has a positive weight on the component. (Each of them includes both positive and negative values in a ratio of 0.5:0.5). PC7 is affected by sector with a positive weight on the component. The sector categories that have the highest contribution to PC7 are industry, technology, TradeService.

By adding the macroeconomic factors to the model, components 1, 2, 9 are obtained as significant, with the regression coefficients $\beta_{PC\#1}$ = 0.16, $\beta_{PC\#2}$ = -0.18, $\beta_{PC\#9}$ = -0.11, and with $R^2_{adj}$ =0.46.

The main factors that contribute to each PC are a relative change in assets, liabilities, cash at the end of a period in PC1, GDP and consumption in PC2, and sector in PC9. Each of these factors has a positive weight in the relevant PC. Combined with the sign of the regression coefficient for each PC, the greater is the relative change in assets, liabilities, cash at the end of a period is, the higher wage increase, but an inverse relationship of wage with a relative change in GDP and consumption. The effect of a sector in PC9 is the same as in the previous model.

For the companies that experienced a decrease in wage, none of the PCs was significant in modelling based on the characteristics of the companies (financial and structural). But, the components 4,8 were significant in significance level of 10%, with $\beta_{PC\#4}$ = 0.03, $\beta_{PC\#8}$ = -0.03, and $R^2_{adj}$ =0.01.

PC4 includes the super-sector and a sector, which have a positive weight on the component, and PC8 includes the profit, with a positive weight as well. The characteristic profit includes positive and negative values in a ratio of 0.5:0.5. The highest contribution of a super-sector to PC4 is of the FinanceTools category, and accordingly the highest contribution of a sector to that PC is of the FinanceService and FinanceTools categories. For PC8, based on its regression coefficient, an increase in the relative change of profit affects the downward change in wage, which means that the wages decreased more. That is, the wage decreases less compared to if the profit changes to negative.

Adding the macroeconomic factors to the model yields that all the principal components are not significant, but components 4, 10 are significant at a significance level of 8%, with regression coefficients $\beta_{PC\#4}$ = -0.03, $\beta_{PC\#10}$ = -0.04, and with $R^2_{adj}$=0.04.

PC4 is affected by super-sector and sector, and PC10 is affected by profit, each of which has a positive weight on the component. Change in profit is a positive one for about 50% of the companies, and vice versa. The highest contribution of a super-sector to PC4 is of the Finance category, and of a sector is of the FinanceService category.

For a positive change in profit, it affects a downward change in wages, meaning that the wage decreases less than if the profit changes negatively.

To summary, the factors that explain a decrease in wage after the switching auditor are a relative increase in profit, another influencing factor is the FinanceService sector. But the percentage explained by these factors is very low - 4% . Economic variables do not influence when taking into account the company's characteristics To.

Table 10. Modelling the increase in wage in the switching year, all companies, years 2007-2012

| Companies with $\Delta W > 0$ (65 companies) | | | | | | | | | | | | | | |
|---|---|---|---|---|---|---|---|---|---|---|---|---|---|---|
| | Without economic factors | | | | | | | | | | | | | |
| | | PC1 | PC2 | PC3 | PC4 | PC5 | PC6 | PC7 | PC8 | PC9 | PC10 | PC11 | PC12 | TOTAL |
| PCA | variance (%) | 15.26 | 10.13 | 9.26 | 8.64 | 7.73 | 6.44 | 5.72 | 5.46 | 5.29 | 4.47 | 4.09 | | 82.48 |
| | Main loadings (weight*) | Assets (0.92) Liabilities (0.92) Cash End (0.92) | Super-Sector (0.72) Sector (0.89) | Super-Sector (0.61) Sector (0.94) | Cash Begin (0.45) Super-Sector (0.45) Sector (0.97) | Super-Sector (0.42) Sector (0.69) | sector (0.55) | Sector (0.79) | Sector (0.73) | Sector (0.49) | Sector (0.41) | Auditor Type (0.20) | | |
| PCR | PCs #1,7 are significant, $\beta_{PC\#1}$**= 0.22, $\beta_{PC\#7}$ = -0.12, $R^2_{adj}$***=0.48 | | | | | | | | | | | | | |
| | With economic factors | | | | | | | | | | | | | |
| PCA | variance (%) | 14.47 | 9.42 | 8.83 | 7.89 | 7.53 | 6.20 | 5.94 | 5.47 | 4.88 | 4.46 | 3.78 | 3.57 | 82.44 |
| | Main loadings (weight) | Assets (0.69) Liabilities (0.69) Cash End (0.69) | GDP (0.49) Consumption (0.40) | Super-Sector (0.69) Sector (0.84) | Super-Sector (0.57) Sector (0.93) | Sector (0.91) Cash Begin (0.45) | Sector (0.49) | sector (0.52) | Sector (0.57) | sector (0.80) | sector (0.52) | Sector (0.40 | sector (0.30) | |
| PCR | PCs #1,2,9 are significant, $\beta_{PC\#1}$ = 0.16, $\beta_{PC\#2}$ = -0.18, $\beta_{PC\#9}$ = -0.11. $R^2_{adj}$=0.46 | | | | | | | | | | | | | |

* The characteristics that have the most influence on the PC are reported, where the correlation of each characteristic with the PC is reported in brackets. ** $\beta_{PC\#k}$ denotes the regression coefficient of the k-th PC *** $R^2_{adj}$ denotes the adjusted-$R^2$.

PCR is the linear regression of the relative difference in wage $\Delta W$ on the PCs.

Table 11. Modelling the decrease in wage in the switching year, all companies, years 2007-2012

| | | PC1 | PC2 | PC3 | PC4 | PC5 | PC6 | PC7 | PC8 | PC9 | PC10 | PC11 | PC12 | TOTAL |
|---|---|---|---|---|---|---|---|---|---|---|---|---|---|---|
| \multicolumn{14}{l}{Companies with $\Delta W < 0$ (78 companies)} | | | | | | | | | | | | | | |
| | \multicolumn{13}{l}{Without economic factors} | | | | | | | | | | | | | |
| PCA | variance (%) | 12.80 | 11.14 | 9.23 | 8.76 | 7.42 | 6.40 | 5.75 | 5.36 | 4.93 | 4.74 | 4.34 | | 80.86 |
| | Main loadings (weight*) | EPS (0.49) Capital (0.48) Cash End (0.82) | Super-Sector (0.48) sector (0.70) | Super-Sector (0.92) Sector (0.93) | Super-Sector (0.97) Sector (0.98) | sector (0.52) | sector (0.40) | Revenue (0.35) | Profit (0.57) | Sector (0.67) | Sector (0.76) | Sector (0.97) | | |
| PCR | \multicolumn{13}{l}{All PCs are insignificant, but PCs #4,8 are significant at 0.095 level, $\beta_{PC\#4}$**= 0.03, $\beta_{PC\#8}$ = -0.03, $R^2_{adj}$***=0.01} | | | | | | | | | | | | | |
| | \multicolumn{13}{l}{With economic factors} | | | | | | | | | | | | | |
| PCA | variance (%) | 11.79 | 9.83 | 9.50 | 8.47 | 7.75 | 6.26 | 5.79 | 4.92 | 4.81 | 4.37 | 4.12 | 3.76 | 81.36 |
| | Main loadings (weight) | CashEnd (0.71) | GDP (0.72) AverageWage (0.45) Sector (0.43) | Super-Sector (0.54) Sector (0.68) Capital (0.4) | Super-Sector (0.68) Sector (0.74) | Super-Sector (0.92) Sector (0.92) | sector (0.42) | sector (0.54) | sector (0.24) | Revenue (0.26) | Profit (0.40) | Sector (0.79) | sector (0.32) | |
| PCR | \multicolumn{13}{l}{All PCs are insignificant, but PCs #4,10 are significant at 0.08 level, $\beta_{PC\#4}$ = -0.03, $\beta_{PC\#10}$ = -0.04, $R^2_{adj}$=0.04} | | | | | | | | | | | | | |

* The characteristics that have the most influence on the PC are reported, where the correlation of each characteristic with the PC is reported in brackets. ** $\beta_{PC\#k}$ denotes the regression coefficient of the k-th PC *** $R^2_{adj}$ denotes the adjusted-$R^2$.

PCR is the linear regression of the relative difference in wage $\Delta W$ on the PCs.

## 4. Results and discussion

In contrast to previous studies in which a decrease in auditor wage was observed in the year of his switching, in this study we found that not all companies experience a decrease in the hourly wage in that year, but in slightly more than half of them. However, it this result has found to be dependent on the period - a year of an economic crisis or a period economic stability. Immediately after an economic crisis, most of the companies experienced an increase in wage in the year of switching compared to the previous year. A decrease in wage occurred at the year of switching when there was an economic crisis, but a period of economic stability. Since not all companies experienced an increase in wage during an economic crisis, the characterization of the economic strength of the economy in a particular year is not sufficient to trace the behavior of the direction of the change in wage. Therefore, the structural characteristics of the companies were first examined as potential influence factors on the direction of the change in wage. We got that no correlation with the audit firm's size, the existence of a going concern note, and the appointment of a senior position in the year of the switching. Likewise, no relationship was accepted between the direction of the wage and the sector, except "Investments and Holdings" sector where a tendency to a decrease in wage was found in the year of the switching, and the biomed sector where a tendency towards an increase in wage was received.

That is, the structural characteristics of the company did not contribute to the explanation of the behavior of the change in wage. Therefore, in the next step, the company's financial characteristics were examined as potential factors influence this behavior. In particular, the relative change of each financial characteristic in the year of switching compared to the previous year was taken. In some companies, larger fluctuations were observed in one or more financial characteristics beyond these two years. Distinguishing between a group of companies with such larger fluctuations ("outlier companies") versus companies with stable fluctuations ("regular companies") brought up the following characteristics of each group: the outlier group was characterized by companies belonging to the real and high-tech super-sector only, while the regular companies also included companies belonging to the financial super-sector. In addition, the outlier companies had a higher percentage of going concern note compared to the regular companies, and a slightly higher percentage of companies that changed the size of audit firm in the year of switching, compared to the regular companies. In terms of the relative difference in wage, these two groups had a similar percentage of companies with an increase in wage (slightly more than half of the companies in each group), however the group of outlier had a higher variability in the size of the wage increase compared to the regular companies.

Because of the difference between the groups, modeling the direction of the change in wage was done for each of them. The modeling included the relative change in the financial characteristics as well as the state of the economy from a macroeconomic point of view. According to this modeling in the outlier companies' group, a relative change in assets, liabilities, and cash at the end of a period has a positive effect on the relative change in wage, meaning a greater relative change in one of them leads to a greater relative change in wage. This is compared to the sector factor that had both a positive and negative effect on the relative difference in wage. The influence of the sector is expressed by tendency to a decrease in wage in the Energy Gas, InvesHold, RealConstruction, and Technology sectors, an increase in wage in the biomed and industry sectors, and an increase/decrease in wage with the same chance in TradeService. These factors together explained 43% of the variance in the relative change in wage.

On the other hand, for the regular companies, the sector was the only factor effected the relative change in wage, while there was no effect of macroeconomic factors. The effect of the sector was manifested by a tendency for decrease in wage in the Energy Gas, Finance Service, Finance Tools, Industry, InvesHold sectors, increase in wage in biomed, RealConstruction , TradeSercice sectors, and an increase/decrease in wage with the same chance in the technology sector. However, the percentage of explanation of the relative change in wages according to this factor was negligible, 4% at most.

Note that while the relationship between the direction of the change in wage and sector was vague for all the companies together, there was an influence of sector on the direction of the change in wage when split the companies into the 2 types of outlier and regular.

The above modeling examines the change in wages in some direction, not necessarily positive or negative. However, it is important also to examine the modeling of a specific direction of the relative change in wage, that is, a positive direction (increase in wage in the year of switching) or negative (a decrease in wage in the year of switching). When referring to all companies together, such modeling produced the following results: a positive direction is explained by a change in assets, liabilities, cash at the end of a period, each of which has a positive effect on wages, by GDP and consumption, each of which has a negative effect on wages, and by sector. The sectors Technology, TradeService, industry had the highest contribution in the effect on wage. These factors together explained 46% of the variation in the relative change in wage.

Negative direction is explained by super-sector and sector, where the highest contribution to explaining the change in wages is by the Finance super-sector, and FinanceService sector. In addition, profit had a negative effect, that is, a positive change in profit leaded to a downward change in wage (the wage decreased less compared to a negative change in profit). However, the percentage of explanation of the change in wage by these factors was very low - 4%, and economic factors had no effect on the change in wage.

An overlap can be seen in terms of the factors influencing an increase in wage beyond all companies and those influencing any change in wage in the outlier companies, as well as between the factors influencing a decrease in wage beyond all companies and those influencing any change in the regular companies. However, this can be expected because, as we found above, the outlier group had a higher variability in the size of the wage increase compared to the regular companies, and therefore the outlier group had a higher influence on the wage modeling in the positive direction (increase in wage).

## 5. Summary

The companies that were considered in this paper are Israeli companies that have switched an auditor in the years 2007-2012. Most of these companies were belong to the real super-sector, mainly in the real construction sector, about 30% of the companies were in the high- tech super-sector, mainly in the "technology" sector, and a minor percentage of the companies in the financial and financial instruments super-sectors.

The change in the hourly wage of the auditor in the switching year compared to the previous year depends on the economic stability of the economy, when the wage is lower in a period of economic stability, but tends to increase in a period of economic depression. In addition, the change in this wage depends on the behavior of the company's financial characteristics, having high variability between the two consecutive years or a stable one. In the former case, the factors affecting the change in the relative wage coincide with those affecting an increase in wage. This is compared to the letter case for which there is an overlap with the factors affecting a decrease in wage. The reason for this result is that the companies with a higher variability in their characteristics had a larger variability in the wage increasing compared to the companies with a stable variability in their characteristics. Therefore, the companies with a higher variability in their characteristics had a higher influence on the wage modeling in the positive direction (that is, increase in wage). According to this, when an auditor and the company manager come to discuss the auditor wage at auditor switching, they should consider two issues: First, the stability of the economy, immediately around a crisis year there is a tendency for higher wages. Second, if it is a company with a large change in its financial characteristics (such companies usually tend to be high-tech companies), a relative change in assets, liabilities, cash, GDP, and consumption should be examined, where the first three characteristics have positive effect on the wage, and the two other factors have a negative effect. In addition, relationship of difference in wage and the sector should be examined.

To predict factors affecting a decrease in wage, or a wage change in a company with stable change in its financial characteristics, the relation of difference in wage and the sector should be examined. But since this relation is small, this issue should be investigated in further research.

**References**

Al- mutairi, A., Naser, K., & Al- Enazi , N. (2017). An empirical investigation of factors affecting audit fees: Evidence from Kuwait. *International advances in economic research*, *23*, 333-347.

Bartlett, MS (1951). The effect of standardization on a χ 2 approximation in factor analysis. *Biometrika* , *38* (3/4), 337-344.

Carcello, J. V., & Neal, T. L. (2003). Audit committee characteristics and auditor dismissals following "new" going-concern reports. The accounting review, 78(1), 95-117.

Castro, WBDL, Peleias, IR, & Silva, GPD (2015). Determinants of audit fees: A study in the companies listed on the BM&FBOVESPA, Brazil. *Revista Contabilidade & Finanças*, *26* , 261-273.

CBS_1. URL: https://www.cbs.gov.il/he/publications/Pages/2008/%D7%9E%D7%93%D7%93-%D7%94%D7%9E%D7%97%D7% 99%D7%A8%D7%99%D7%9D-%D7%9C%D7%A6%D7%A8%D7%9B%D7%9F-%D7%93%D7%A6%D7%9E%D7%91%D7%A8-2008.aspx

CBS_2. URL: https://www.cbs.gov.il/he/subjects/Pages/%D7%A9%D7%9B%D7%A8-%D7%94%D7%9B%D7%A0%D7%A1% D7%95%D7%AA.aspx

CBS_3. URL: https://www.cbs.gov.il/he/subjects/Pages/%D7%A9%D7%9B%D7%A8-%D7%94%D7%9B%D7%A0%D7%A1%D7% 95%D7%AA.aspx

CBS_4. URL: https://www.cbs.gov.il/he/subjects/Pages/%D7%A9%D7%9B%D7%A8-%D7%94%D7%9B%D7%A0%D7%A1% D7%95%D7%AA.aspx

Companies Law, 1999, Section 165.

DeAngelo, LE (1981). Auditor independence, 'low balling', and disclosure regulation. Journal of accounting and economics, 3(2), 113-127.

Firth, M. (1999). Company takeovers and the auditor choice decision. Journal of International Accounting, Auditing and Taxation, 8(2), 197-214.

Greenacre, M., Groenen, PJ, Hastie, T., d'Enza, AI, Markos, A., & Tuzhilina, E. (2022). Principal component analysis. *Nature Reviews Methods Primers*, *2* (1), 100.

Harjoto, MA, Laksmana, I., & Lee, R. (2015). The impact of demographic characteristics of CEOs and directors on audit fees and audit delay. *Managerial Auditing Journal*.

Hennes, K. M., Leone, A. J., & Miller, B. P. (2010). Accounting restatements and auditor accountability. Unpublished manuscript, University of Oklahoma, United State.

Historical rates. URL: https://www.poundsterlinglive.com/bank-of-england-spot/historical-spot-exchange-  .rates/ usd /USD-to-ILS-2012 .

Horowitz, A., Eden, Y., & Moore, D. R. (2017). The factors affecting the amount of fees of the accountants of the public companies in Israel. The Economic Quarterly, 61(3/4), 103-118. (Hebrew).

Jolliffe, IT (2010). Principal components in regression analysis. *Principal Component Analysis; Jolliffe, IT, Ed.; Springer Series in Statistics*, 129-155.

Johnson, W. B., & Lys, T. (1990). The market for audit services: Evidence from voluntary auditor changes. Journal of accounting and economics, 12(1-3), 281-308.

Kaiser, HF, & Rice, J. (1974). Little jiffy, mark IV. *Educational and psychological measurement*, *34* (1), 111-117.

Kikhia, HY (2015). Determinants of audit fees: Evidence from Jordan. *Accounting and Finance Research*, *4* (1), 42-53.

Kimeli, EK (2016). Determinants of audit fees pricing: Evidence from Nairobi Securities Exchange (NSE).

Liu, S. (2017). An empirical study: Auditors' characteristics and audit fees. *Open Journal of Accounting*, *6* (2), 52-70.

Magna Web. URL: https://www.magna.isa.gov.il/default.aspx

Maya web. URL: https://maya.tase.co.il/reports/details/1216511/2/0

Mohammed, NH, & Saeed, A. (2018). Determinants of audit fees: Evidence from the UK alternative investment market. *Academic Journal of Nawroz University*, *7* (3), 34-47.

Musah, A. (2017). Determinants of audit fees in a developing economy: Evidence from Ghana. *International Journal of Academic Research in Business and Social Sciences*, *7* (11), 716-730.

Naser, K., & Hassan, YM (2016). Factors influencing external audit fees of companies listed on Dubai Financial Market. *International Journal of Islamic and Middle Eastern Finance and Management*.

Simon, D. T., & Francis, J. R. (1988). The effects of auditor change on audit fees: Tests of price cutting and price recovery. Accounting Review, 255-269.

Super, SO, & Shil, NC (2019). Determinants of Audit Fee in the Manufacturing Sector in Nigeria. *IUP Journal of Accounting Research & Audit Practices*, *18* (2).

Urhoghide, RO, & Izedonmi, FOI (2015). An empirical investigation of audit fee determinants in Nigeria. *International Journal of Business and Social Research*, *5* (8), 48-58.

Weiss, D., Bezalel, A., Moisa, E., Nistel, S., Friedman, A. (2021). Does a replacement of the accounting firm affect audit fees? innovations in management, 9, 126-134. (In Hebrew). URL: https://coller.tau.ac.il/sites/coller.tau.ac.il/files/media_server/Recanati/management/newsletter/august2021/Weiss.pdf

# Appendices

## Appendix A: Correlations, all companies, years 2007-2012

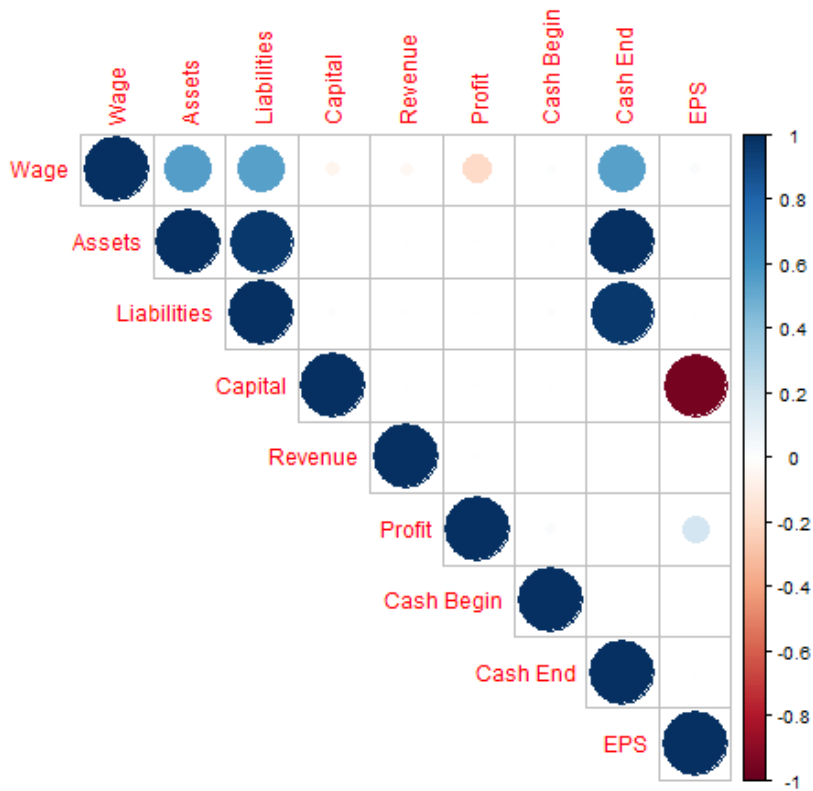

**Figure A1.** Correlation matrix of relative change in each characteristic financial and relative change in wage, all companies, years 2007-2012.

**Table A1.** Correlations between relative change in financial characteristic and a relative change in wage, all companies together. The correlations are described at each year, and in the All row over all the years 2007-2012 together.

| Year | Assets | Liabilities | Capital | Revenue | Profit | Cash Begin | Cash End | EPS |
|---|---|---|---|---|---|---|---|---|
| 2007 | 0.88 | 0.88 | -0.88 | -0.12 | -0.05 | -0.19 | 0.88 | -0.07 |
| 2008 | -0.10 | 0.27 | 0.12 | -0.08 | -0.40 | 0.08 | -0.10 | 0.21 |
| 2009 | 0.69 | 0.02 | -0.06 | -0.42 | 0.27 | -0.05 | 0.30 | 0.05 |
| 2010 | -0.10 | 0.07 | -0.22 | 0.26 | 0.03 | -0.11 | 0.11 | 0.09 |
| 2011 | 0.01 | 0.04 | -0.07 | 0.01 | -0.31 | 0.30 | -0.11 | -0.32 |
| 2012 | -0.18 | -0.35 | -0.15 | -0.21 | 0.02 | -0.03 | -0.24 | 0.17 |
| All | 0.55 | 0.54 | -0.05 | -0.04 | -0.20 | 0.02 | 0.55 | 0.03 |

**Table A2.** Joint distribution (relative frequency) of sector according to direction of relative difference in wage $\Delta W$. All companies, years 2007-2012.

| Sector | $\Delta W <0$ | $\Delta W >0$ |
|---|---|---|
| Biomed | 0.03 | 0.06 |
| Energy | 0.03 | 0.01 |
| Financial Services | 0.02 | 0.01 |
| Financial Instruments | 0.02 | 0.01 |
| Industry | 0.05 | 0.06 |
| Investment | 0.11 | 0.06 |
| Real Estate | 0.12 | 0.11 |
| Technology | 0.10 | 0.08 |
| Trade & Services | 0.06 | 0.06 |

## Appendix B. Outlier companies (having large variability in their characteristics)

**Table B1.** Distribution of outlier companies (relative frequency) by type of super-sector and sector

| Super-Sector | Number of companies | High-Tech | | | Finance | Real | | | | | | Financial Instruments |
|---|---|---|---|---|---|---|---|---|---|---|---|---|
| Sector | | Biomed | Technology | ALL | FinanceServ | TradeServ | Real-Estate | Industry | Investment | Energy | ALL | FinanceInstru |
| year | | | | | | | | | | | | |
| 2007 | 11 | 0.00 | 0.27 | 0.27 | 0.00 | 0.00 | 0.45 | 0.18 | 0.00 | 0.09 | 0.73 | 0.00 |
| 2008 | 14 | 0.21 | 0.00 | 0.21 | 0.00 | 0.14 | 0.36 | 0.07 | 0.07 | 0.00 | 0.79 | 0.00 |
| 2009 | 10 | 0.1 | 0.1 | 0.2 | 0.00 | 0.00 | 0.40 | 0.20 | 0.20 | 0.00 | 0.80 | 0.00 |
| 2010 | 9 | 0.33 | 0.11 | 0.44 | 0.00 | 0.11 | 0.11 | 0.00 | 0.33 | 0.00 | 0.56 | 0.00 |
| 2011 | 10 | 0.0 | 0.2 | 0.2 | 0.00 | 0.10 | 0.20 | 0.00 | 0.30 | 0.20 | 0.80 | 0.00 |
| 2012 | 10 | 0.3 | 0.3 | 0.6 | 0.00 | 0.00 | 0.30 | 0.00 | 0.00 | 0.10 | 0.40 | 0.00 |
| **ALL** | **64** | **0.16** | **0.16** | **0.31** | **0.00** | **0.06** | **0.31** | **0.08** | **0.17** | **0.06** | **0.69** | **0.00** |

.

**Table B2.** Distribution of outlier companies (relative frequency) by type of audit firm change, per year

|       | Audit firm change |               |              |
|-------|-------------------|---------------|--------------|
| year  | No difference     | Small to big5 | Big5 to small |
| 2007  | 0.45              | 0.36          | 0.18         |
| 2008  | 0.79              | 0.07          | 0.14         |
| 2009  | 0.8               | 0.1           | 0.1          |
| 20 10 | 0.56              | 0.22          | 0.22         |
| 2011  | 0.6               | 0.4           |              |
| 2012  | 0.5               | 0.2           | 0.3          |
| **ALL** | **0.62**        | **0.22**      | **0.16**     |

**Table B3.** Distribution of outlier companies (relative frequency) per year according to existence of a going concern note and a senior position appointment, at the switching year.

| year | Going Concern | Senior Position |
|------|---------------|-----------------|
| 2007 | 0.18          | 0.73            |
| 2008 | 0.57          | 0.86            |
| 2009 | 0.5           | 0.7             |
| 2010 | 0.33          | 0.78            |
| 2011 | 0.4           | 0.7             |
| 2012 | 0.6           | 0.8             |
| **ALL** | **0.44**   | **0.77**        |

**Table B4.** Distribution of outlier companies (relative frequency) per year according to the direction of the relative difference in wage, $\Delta W$.

| Year | $\Delta W < 0$ | $\Delta W > 0$ |
|------|----------------|----------------|
| 2007 | 0.36           | 0.64           |
| 2008 | 0.64           | 0.36           |
| 2009 | 0.3            | 0.7            |
| 2010 | 0.56           | 0.44           |
| 2011 | 0.7            | 0.3            |
| 2012 | 0.6            | 0.4            |
| **ALL** | **0.53**    | **0.47**       |

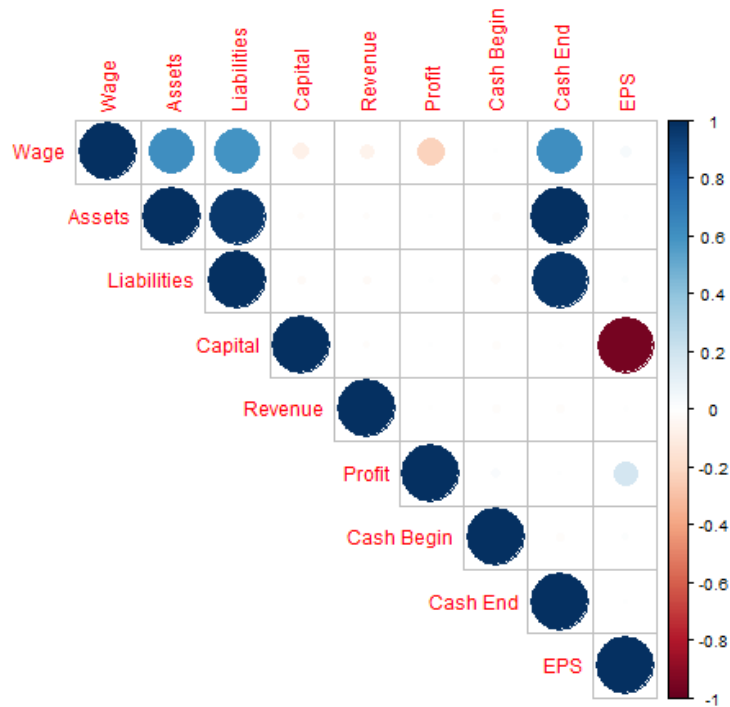

**Figure B1.** Correlations among the financial characteristics of outlier companies, years 2007-2012.

**Appendix C.** Regular companies (having small variability in their characteristics)

Table C1. Distribution of regular companies (relative frequency) by type of super-sector and sector

| Super-Sector | Number of companies | High-Tech | | | Finance | | Real | | | | | | Financial Instruments |
|---|---|---|---|---|---|---|---|---|---|---|---|---|---|
| Sector | | Biomed | Technology | ALL | FinanceServ | TradeServ | Real-Estate | Industry | Investment | Energy | ALL | | FinanceInstru |
| year | | | | | | | | | | | | | |
| 2007 | 9 | 0.00 | 0.33 | 0.33 | 0.11 | 0.00 | 0.11 | 0.22 | 0.22 | 0.00 | 0.56 | | 0.00 |
| 2008 | 25 | 0.00 | 0.16 | 0.16 | 0.00 | 0.36 | 0.24 | 0.08 | 0.16 | 0.00 | 0.84 | | 0.00 |
| 2009 | 10 | 0.10 | 0.20 | 0.30 | 0.10 | 0.10 | 0.10 | 0.20 | 0.20 | 0.00 | 0.60 | | 0.00 |
| 2010 | 15 | 0.07 | 0.20 | 0.27 | 0.07 | 0.00 | 0.13 | 0.13 | 0.27 | 0.07 | 0.60 | | 0.07 |
| 2011 | 10 | 0.00 | 0.20 | 0.20 | 0.00 | 0.20 | 0.10 | 0.10 | 0.10 | 0.10 | 0.60 | | 0.20 |
| 2012 | 10 | 0.00 | 0.20 | 0.20 | 0.20 | 0.10 | 0.20 | 0.20 | 0.00 | 0.00 | 0.50 | | 0.10 |
| **ALL** | **79** | **0.03** | **0.20** | **0.23** | **0.06** | **0.16** | **0.16** | **0.14** | **0.16** | **0.03** | **0.66** | | **0.05** |

.

**Table C2.** Distribution of regular companies (relative frequency) by type of audit firm change, per year

|  | **Audit firm change** | | |
|---|---|---|---|
| Year | No difference | Small to big5 | Big5 to small |
| 2007 | 0.33 | 0.44 | 0.22 |
| 2008 | 0.44 | 0.32 | 0.24 |
| 2009 | 0.5 | 0.1 | 0.4 |
| 2010 | 0.47 | 0.33 | 0.20 |
| 2011 | 0.4 | 0.4 | 0.2 |
| 2012 | 0.6 | 0.2 | 0.2 |
| **ALL** | **0.46** | **0.30** | **0.24** |

**Table C3.** Distribution of regular companies (relative frequency) per year according to existence of a going concern note and a senior position appointment, at the switching year.

| Year | Going Concern | Senior Position |
|---|---|---|
| 2007 | 0 | 0.67 |
| 2008 | 0.2 | 0.76 |
| 2009 | 0.3 | 0.8 |
| 2010 | 0.2 | 0.8 |
| 2011 | 0.1 | 0.8 |
| 2012 | 0.3 | 0.9 |
| **ALL** | **0.19** | **0.78** |

**Table C4.** Distribution of regular companies (relative frequency) per year according to the direction of the relative difference in wage, $\Delta W$.

| Year | $\Delta W < 0$ | $\Delta W > 0$ |
|---|---|---|
| 2007 | 0.56 | 0.44 |
| 2008 | 0.56 | 0.44 |
| 2009 | 0.3 | 0.7 |
| 2010 | 0.53 | 0.47 |
| 2011 | 0.6 | 0.4 |
| 2012 | 0.8 | 0.2 |
| **ALL** | **0.56** | **0.44** |

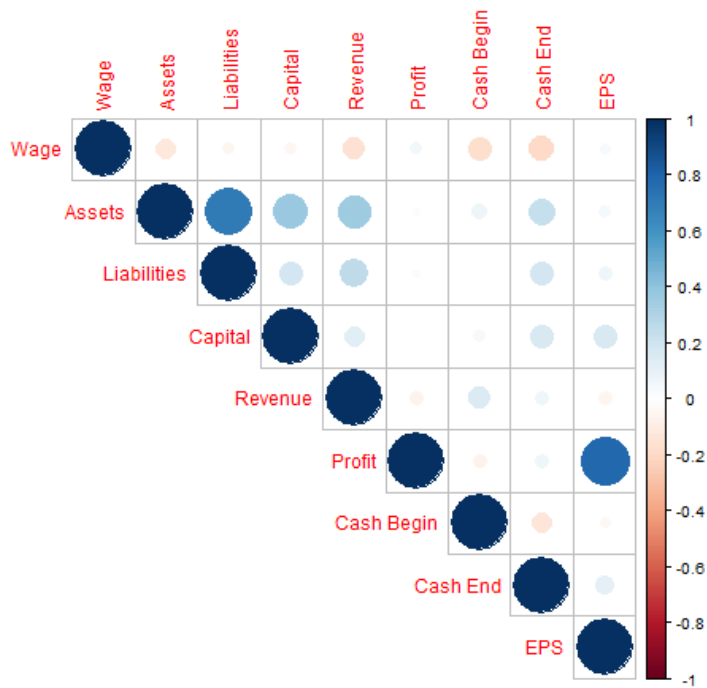

**Figure C1. Correlations among the financial characteristics of the regular companies, years 2007-2012.**

**Appendix D. VIF index For multicollinearity, all companies, years 2007-2012.**

| feature | EPS | Assets | Liabilities | Capital | Revenue | Profit | Cash Begin | Cash End |
|---|---|---|---|---|---|---|---|---|
| VIF | 18.81 | 4882.60 | 18.55 | 18.61 | 1.04 | 1.70 | 1.03 | 5045.30 |
| feature | Price | GDP | Consumption | Market Wage | | | | |
| VIF | 10.19 | 40.26 | 30.02 | 20.29 | | | | |

**Appendix E. PCR model for all companies per year**

**Table E1. Goodness of fit of the Data (company's characteristics) to PCA, per year**

| Year | Bartlett p-value | KMO Overall |
|---|---|---|
| 2007 | 2.50E-152 | 0.48 |
| 2008 | 2.36E-72 | 0.75 |
| 2009 | 0.99 | 0.52 |
| 2010 | 3.77E-19 | 0.47 |
| 2011 | 2.49E-61 | 0.55 |
| 2012 | 9.46E-81 | 0.43 |

**Table E2. PCR of relative difference in wage on principal components (PCs) obtained by PCA, per year.**

| Year | PCA | | | | |
|---|---|---|---|---|---|
| **2007** | **6 PCs, V=79.11** | | | | |
| | **PC1** (22.93) | Assets | Liabilities | Capital | Cash End |
| | **PC2** (14.46) | Super-Sector | Sector | | |
| | **PC3** (12.28) | EPS | Profit | | |
| | **PC4** (11.63) | Super-Sector | Sector | | |
| | **PC5** (9.97) | Sector | | | |
| | **PC6** (7.84) | Sector | | | |
| **PCR** | PC #1 is significant, $R^2_{adj}$=0.76 | | | | |
| **2008** | **8 PCs, V=82.55** | | | | |
| | **PC1** (22.74) | Assets | Capital | Revenue | Cash End |
| | **PC2** (13.95) | Sector | | | |
| | **PC3** (11.75) | Sector | Super-Sector | | |
| | **PC4** (9.49) | Sector | Profit | | |
| | **PC5** (7.06) | Sector | Senior Position | | |
| | **PC6** (6.25) | Auditor Type | | | |
| | **PC7** (6.13) | Sector | | | |
| | **PC8** (5.18) | Sector | | | |
| **PCR** | All PCs are insignificant, except PC #4 which is significant at 8%, $R^2_{adj}$=0.06 | | | | |
| **2009** | **8 PCs, V= 81.40** | | | | |
| | **PC1** (16.38) | Sector | Super-Sector | | |
| | **PC2** (14.34) | Sector | | | |
| | **PC3** (12.44) | Sector | Auditor Type | | |
| | **PC4** (10.40) | Sector | | | |
| | **PC5** (8.62) | Sector | | | |
| | **PC6** (7.31) | Sector | | | |
| | **PC7** (6.55) | Cash Begin | Going Concern | | |
| | **PC8** (5.35) | Sector | | | |
| **PCR** | All PCs are insignificant, except PC #2 which is significant at 8%, $R^2_{adj}$=0.14 | | | | |
| **2010** | **8 PCs, V= 79.57** | | | | |

|      | PC1 (20.18) | Profit | Sector | Capital | EPS |
|------|-------------|--------|--------|---------|-----|
|      | PC2 (12.7)  | Sector |        |         |     |
|      | PC3 (10.17) | Sector | Super-Sector |   |     |
|      | PC4 (9.42)  | Sector | Super-Sector |   |     |
|      | PC5 (8.80)  | Sector |        |         |     |
|      | PC6 (7.08)  | Sector |        |         |     |
|      | PC7 (5.77)  | Sector |        |         |     |
|      | PC8 (5.46)  | Sector |        |         |     |
| PCR  | All PCs are insignificant ||||||
| 2011 | 7 PCs, V= 81.92 ||||||
|      | PC1 (24.36) | Assets | Liabilities | Revenue | Capital |
|      | PC2 (13.91) | EPS    | Profit      | Sector  |         |
|      | PC3 (12.65) | Super-Sector | Sector |       |         |
|      | PC4 (9.38)  | Sector | Senior Position | Going Concern | |
|      | PC5 (8.14)  | Sector |        |         |     |
|      | PC6 (6.85)  | Sector |        |         |     |
|      | PC7 (6.64)  | Sector |        |         |     |
| PCR  | All PCs are insignificant, except PC #2 which is significant at 5%, $R^2_{adj}$=0.15 |||||
| 2012 | 7 PCs, V= 82.22 ||||||
|      | PC1 (21.62) | Cash End | Revenue | Assets | Profit |
|      | PC2 (18.08) | Sector   | EPS     | Capital |        |
|      | PC3 (10.63) | Sector   |         |         |        |
|      | PC4 (10.17) | Sector   | Super-Sector |   |        |
|      | PC5 (9.57)  | Sector   | Super-Sector |   |        |
|      | PC6 (6.39)  | Sector   |         |         |        |
|      | PC7 (5.76)  | Sector   |         |         |        |
| PCR  | All PCs are insignificant ||||||

The PCA is based on financial and structural characteristics of the companies.

The PCA results include: (i) the number of PCs that obtained a cumulative explained variance of approximately 80% (ii) the percentage of cumulative explained variance V based on the considered PCs (iii) the percentage of variance explained by each PC is presented in parentheses (iv) the company characteristics which have a high impact on each PC are listed.

The PCR results include the significant PCs and the resulted adjusted R^2 denoted by $R^2_{adj}$.